\documentclass[12pt] {article}

\usepackage{amsmath}

\begin{document}

\begin{center}
{\large \bf On the classical treatment of the Stark effect in 
hydrogen atoms}
\end{center}

\vskip 0.5cm
\begin{center}
{\bf Fabio Sattin}\footnote{
{\it Present address}: 
Consorzio RFX,
Corso Stati Uniti 4, 35127 Padova, Italy. E--mail: 
sattin@igi.pd.cnr.it} 

\vskip 0.3cm

{\em Dipartimento di Ingegneria Elettrica, Universit\`a di Padova, \\
Via Gradenigo 6/a, 35131 Padova, Italy  \\
Istituto Nazionale di Fisica della Materia, Unit\`a di Padova, \\
Corso Stati Uniti 4, 35127 Padova, Italy }
\end{center}
\vskip 0.7cm
{\parindent = 0.pt
{\bf Summary.} --- A classical model of the hydrogen atom in a
static electric field is studied, basing upon the work
[ Hooker A. {\it et al}, {\it Phys. Rev. A}, {\bf 55} (1997) 4609 ].
In that work the electrons are supposed to move along Kepler orbits
around the nucleus, while interacting with the external field.
That classical model reproduces very well the true energy shift of 
the linear Stark effect. The agreement with the second order effect is 
poor. It is shown here that the results for the quadratic Stark 
effect may be considerably improved if the electrons are still allowed to move
along classical trajectories, but whose initial conditions are 
statistically sampled from a distribution resembling the quantum 
mechanical
one.

aspects
 }

\newpage

\section*{1. - Introduction}

The use of classical models to study quantum mechanical systems is 
largely employed in microscopic physics. For large quantum numbers it is 
justified on the basis of the correspondence principle. Also at small quantum 
numbers its utility is not negligible since it allows at least to obtain a 
qualitative insight of the dynamics of the system; in the best cases, it is even 
possible to extract quantitative results in a easy way and with a good 
accuracy. 
Some examples are the scattering of electrons from an atom 
\cite{elettrone1, elettrone2}, and the treatment of energetic 
ion--atom collisions \cite{ione1, ione2, ione3}. 
Classical model are obviously favoured over their quantum--mechanical
counterparts from the {\em numerical} point of view. In those cases where 
also the quantum problem is solvable, the classical picture is important  
since it allows to  clearly visualize the problem at hand. For these 
reasons the correspondence between the classical and quantum mechanical 
descriptions is currently a well studied topic. 

Quite recently it has been proposed  a purely classical view to the Stark effect 
in hydrogen \cite{starkclassico}: the field is supposed to interact with classical 
electrons following Kepler orbits around the nucleus. Following
this work, in ref. \cite{hooker97} the energy shifts for the linear 
and quadratic Stark effect have been computed. The agreement between the classical
result and the exact quantum mechanical one is excellent for the linear effect
but is not good for the quadratic effect, which only asymptotically for
large quantum numbers approaches the correct value.

The purpose of the present paper is to show that, within a purely classical 
formalism\footnote{ It must be stressed that any ''classical'' model 
necessarily incorporates some elements of the quantum theory in order to 
simulate a microscopic object. 
The relevance of the approach \cite{hooker97} and of the present paper is that 
only initial conditions are to be chosen compatible with the laws of 
quantum mechanics.}
very similar to that of \cite{hooker97}, the quadratic Stark effect may 
be reproduced very accurately: it is enough to relax the bounds on the electron 
trajectory and adopt a statistical approach:  here, the electron is still modelled 
as a classical particle following Kepler orbits but the initial conditions are 
picked up from a distribution subject to certain rules.  It will be shown that 
a much better agreement is obtained by this slightly more sophisticated approach.

Atomic units will be used throughout this work unless explicitly 
stated.

\section*{2. - The classical model of \cite{hooker97}}

As well known, the study of the Stark effect is easily performed in 
parabolic coordinates, where the hydrogen atom is classified by the quantum
numbers $ n , n_1 , n_2 , m $. We are considering small fields,
for which a perturbative approach is adequate and they still are 
good quantum numbers.

I will first briefly summarize the treatment of the first--order 
Stark effect, which will be useful also to introduce the main concepts used later. 
First of all, let us recall some results stated in 
\cite{starkclassico}. In that work a  correspondence has been done between the 
quantum mechanical operators and their classical counterparts:
\begin{itemize}
\item to each set of values $ n_1, n_2, n, m$ there corresponds a set 
of classical Keplerian  orbits of the electron which are generally  elliptical, 
with eccentricity
$ \varepsilon = \sqrt{ 1 - l^2/n^2}$ ($l$ is angular quantum number 
in spherical coordinates, and the classical angular momentum). When 
time--averaged  over a period of rotation, the ellipse yields a non zero mean 
electric dipole moment: 
\begin{equation}
 <d> = {1 \over T} \int_0^T z \, dt =   3/2 \, n^2 \varepsilon \quad .
\end{equation}

\item The Lenz vector $ {\bf A} = {\bf p \times L} - {\bf r/|r|}$ is 
classically a conserved quantity, as is its $z$--component in quantum 
mechanics: 
\begin{equation}
\label{eq:lenzz}
 A_z = ( n_1 - n_2) n \quad .  
\end{equation}
$A$ is related to the electron orbit by $|{\bf A}| = \varepsilon$. 
Since, besides this, in \cite{starkclassico} it was shown that ${\bf d}$ and 
${\bf A}$ point towards the same direction, one may identify them 
through
\begin{equation}
\label{eq:dipolo}
{\bf d} = (3/2) n^2 {\bf A} \quad .
\end{equation}

\item The energy shift in presence of an electric field ${\bf F}$ is 
calculated from
\begin{equation}
\label{eq:shift1ordine}
\Delta E^{(1)} = - {\bf d} \cdot {\bf F} \quad .
\end{equation}
It is immediate to see that this result agrees with the energy shift
obtained using quantum mechanics \cite{qm91} provided we use for $d$ 
its form
(\ref{eq:dipolo}) with $A$ given by (\ref{eq:lenzz}):
\begin{equation}
\label{eq:shift1ordine1}
\Delta E^{(1)} = - {3 \over 2} n (n_1 - n_2) |F| \qquad ,
\end{equation}
\end{itemize}

In \cite{hooker97} it was verified by numerically integrating the 
Kepler orbits in presence of an electric field  and with initial 
conditions compatible with the quantized value (\ref{eq:lenzz}) that 
the classical energy shifts from its unperturbed value of the 
quantity (\ref{eq:shift1ordine1}).
 
The quadratic effect is of relevance when $ n_1 = n_2$, in which case 
the linear term vanishes. This means--in the language of 
\cite{starkclassico}--that the mean dipole moment vanishes, and this 
happens when the orbits are circular and lying on a plane 
perpendicular to the field axis. 
The electric field may induce a dipole moment by shifting  the 
electron and the nucleus with respect to each other. The energy shift 
for the quadratic Stark effect is defined by
\begin{equation}
\Delta E^{(2)} \equiv - {1 \over 2} \alpha F^2 \quad,
\end{equation}
$ \alpha $ being the polarizability. $ \alpha $ is related to the 
induced dipole moment by 
\begin{equation}
\label{eq:alfadefinizione}
 \alpha {\bf F} = {\bf d} \quad . 
\end{equation}
The quantum mechanical value of $ \alpha $ for states with $ n_1 = 
n_2$ and 
$ m = n - 1$ is \cite{qm91}
\begin{equation}
\label{eq:alfaquantica}
\alpha_{QM} = { n^4 \over 4} ( 4 n^2 + 9 n + 5) \quad .
\end{equation}
In \cite{hooker97} a simple approximation is used to compute the classical 
value of $\alpha $, $ \alpha_{cl}$: be $r$ the radius
of the circular orbit of the electron, and $ \delta z$ the shift along
the direction of {\bf F} between the electron and the nucleus induced 
by the external field. For small values of $ \delta z$ we may approximate 
the Coulomb force on the electron as $ \delta z /r^3$, which balances the force 
exerted on the electron by {\bf F} when
\begin{equation}
\label{eq:forza}
F = { \delta z \over r^3 }  \quad .
\end{equation}
The induced average dipole moment is $ d = \delta z$ and, from eq. (\ref{eq:forza}) 
and the definition of $\alpha $ (\ref{eq:alfadefinizione}),
\begin{equation}
\label{eq:alfaclassica}
\alpha_{cl} = r^3 = n^6
\end{equation}
since for circular Bohr orbits $ r = n^2$.

Even if this is a very simplified model, it was verified in 
\cite{hooker97} that it describes very accurately the classical system: the 
Hamilton's equations for the electron were numerically solved in presence of 
the electric field. 
The electron energy was determined from its position and momentum. 
The numerical results were found to agree well with eq. 
(\ref{eq:alfaclassica}) (see fig. 3 of \cite{hooker97}). It is 
therefore correct to assume that $ \alpha_{cl}$ is well approximated 
by eq. (\ref{eq:alfaclassica}). 

Eqns. (\ref{eq:alfaquantica}) and (\ref{eq:alfaclassica}) only agree 
for $ n \to \infty $; the greater discrepancies are for small $n$s: for 
example, when $n = 1$  the ratio is $ \alpha_{cl}/\alpha_{QM} = 2/9 \approx 
0.22$. 

\section*{3. - Improvements over the simple model}

Both quantum and classical mechanics admit a remarkable unified 
description in terms of fluid dynamics: the Schr\"odinger equation
\begin{equation}
\label{eq:schrodinger}
i \hbar { \partial \psi (r, t) \over \partial r} = \left( - { \hbar^2 
\over 2 m} 
\nabla^2 + U(r, t) \right) \psi( r, t)
\end{equation}
through the replacement
\begin{equation}
\psi(r, t) = \sqrt{\rho(r, t)} \exp\left( { i \over \hbar} S(r, t) 
\right)
\end{equation}
may be rewritten  into the set of two equations
\begin{align}
\label{eq:fluidoquantico}
&{\partial \rho \over \partial t} + \nabla \cdot (\rho {\bf v}) = 0 \\
&m {d {\bf v} \over dt} = - \nabla U_{eff}  
\end{align}
 where
\begin{eqnarray}
U_{eff} &=& U(r, t) - {\hbar^2 \over 2 m} {\nabla^2 \sqrt{\rho} \over 
\sqrt{\rho}} = U(r, t) + U_q(r, t) \\
 {\bf v}(r, t) &=& (\nabla S)/m
\end{eqnarray}

The time evolution of a classical phase space distribution is given 
by the Liouville equation
\begin{eqnarray}
{\partial f \over \partial t} &=& L(t) f \\
L(t) &=& \nabla_r U \cdot \nabla_p - {{\bf p} \over m} \cdot \nabla_r 
\end{eqnarray}
from which, multiplying by 1 and ${\bf p}$ respectively and then 
integrating,
\begin{align}
\label{eq:fluidoclassico}
&{\partial \rho \over \partial t} + \nabla \cdot (\rho {\bf v}) = 0 \\
&m{\partial d {\bf v} \over dt} = - \nabla_r U - {1 \over \rho} 
\nabla_r \Pi \\
&\rho(r, t) = \int d{\bf p} \, f({\bf r},{\bf p},t) \\
& {\bf v} = { 1 \over m \rho} \int d{\bf p} {\bf p} f({\bf r},{\bf p}, t) \\
& \Pi_{ij} = { 1 \over m} \int d{\bf p} p_j p_j f({\bf r},{\bf p}, t) - \rho v_i v_j
\end{align}
The only difference between the two sets of equations is that the 
role of the quantum potential  $U_q$ has been replaced by the stress 
tensor $\Pi$.

From $f$ one gets the projection over the position and momentum 
coordinates
\begin{equation}
\label{eq:proiezioner}
\rho({\bf r}, t) = \int d {\bf p} f({\bf r},{\bf p}, t)  
\end{equation}
\begin{equation}
\label{eq:proiezionep}
\tilde{\rho}({\bf p}, t) = \int d {\bf r} f({\bf r},{\bf p}, t)
\end{equation}

In classical approximations the quantum dynamics of a system is 
computed through an averaging over an ensemble of classical trajectories. 
The average is done over the statistical distribution of initial 
conditions $f$, with $f$ chosen such as to closely reproduce the wave 
function in position or momentum space.

If $f$ is to be stationary (as is the case here), it
can depend only on constants of the motion \cite{cohen85}. 
A largely employed choice is to make $f$ depending only upon the 
energy: $ f ({\bf r},{\bf p}, t) \equiv f(E) $. There is not an unique choice
possible for $f$ since--classically--does not exist any $f$ such that 
both $ \rho $ and $ \tilde{\rho}$ are equal to their quantum 
mechanical counterpart. The choice is done on the basis of mathematical 
simplicity and accuracy. 

Once a choice for $f$--and therefore for $ \rho, \tilde{\rho}$--is 
done, the calculations of the previous section still hold: simply, they 
are to be repeated for an ensemble 
of electrons, {\it each} of them is still obeying Newton's equations, 
and in particular follows a circular orbit. Therefore, eq. 
(\ref{eq:alfaclassica}) is replaced by its average
\begin{equation}
\label{eq:alfamedia}
\alpha = \, < r^3 > \, = \int d {\bf r} \, r^3 \rho({\bf r}) \quad .
\end{equation}
Notice that the model of \cite{starkclassico, hooker97} may be seen 
as a particular case, by putting $ \rho \sim \delta( {\bf r}(t) - {\bf 
r}_K(t)), \tilde{\rho} \sim \delta( {\bf p}(t) - {\bf p}_K(t))$, with 
${\bf r}_K, {\bf p}_K$ position and momentum corresponding to a Kepler orbit. 

I will consider three special cases: 
\begin{enumerate}
\item first of all, the true quantum mechanical distribution is used 
for $ \rho $ 
\begin{equation} 
\label{eq:rhoquantica}
\rho({\bf r}) = |\psi_{n l m}({\bf r})|^2  
\end{equation}
with $ l = |m| =  n - 1 $.
This is an obvious choice, and is done in order to give an insight of
the effectiveness of the method when applied under those which should 
be the conditions closest to the true ones. It has the defect that $ \rho $ now 
encompasses  a region classically forbidden to the electron: from
\begin{equation}
\label{eq:energiaclassica}
{ p^2 \over 2} - { 1 \over r} = - { 1 \over 2 n^2} 
\end{equation}
one gets $ r \leq 2 n^2 $.
\item The second choice is therefore to use a ``truncated'' 
distribution:
\begin{equation}
\label{eq:rhotroncata}
\rho = \begin{cases} C |\psi_{n l m}(r)|^2&  \quad , r < 2 n^2  \\
      0& \quad, r \geq 2 n^2
     \end{cases}
\end{equation}
where $C$ stands for a normalization constant.
\item Finally, I use a microcanonical distribution
\begin{equation}
\label{eq:fmicrocanonica}
f(E)  = { 1 \over 8 \pi^3} \delta\left( E + { 1 \over 2 n^2} \right)
\end{equation}
from which one obtains, after substitution in eq. (\ref{eq:proiezioner}),
\begin{equation}
\label{eq:rhomicrocanonica}
\rho \propto \left( { 1 \over r} - { 1 \over 2 n^2} \right)^{1/2}
\end{equation}
which is a real quantity only for $ r \leq 2 n^2 $. This microcanonical
distribution is largely used in classical calculations of ion--hydrogen 
scattering processes for generating initial electron distributions, 
since it has the property of correctly reproducing the true electron 
momentum distribution $ \tilde{\rho}$ \cite{cohen85}.
\end{enumerate}

All the integrals (\ref{eq:alfamedia}) may be performed analytically 
using the $\rho$s of items 1-3. In table 1 I report the 
values for the polarizability obtained using 1--3, together with the 
values from eqns. (\ref{eq:alfaquantica}) and (\ref{eq:alfaclassica}). It is 
clearly discernible the slow convergence of $ \alpha_{cl}$ to $ \alpha_{QM}$ 
if compared, in particular, to $ \alpha_{2}$ which already at $ n = 1$ is 
within $ 20 \% $ of the true value.
It is curious, on the other hand, the behaviour of $ \alpha_{3}$ 
which is always very close to $ \alpha_{QM}$ and, for $ n = 4 , 5$ is
the best approximation, but becomes worse 
at large $n$s. Another point to remark is that, for large $n$, $\alpha_1$
and $ \alpha_2$ become nearly equal, as it should be from their 
definition.
On the average, using any of the suggested distributions allows at 
least to halve the error with respect to the results of ref. \cite{hooker97}.

\section*{4. - Conclusions}
In ref. \cite{hooker97} it is remarked that the main motivation of 
their work was to show that some quantal results may be recovered, at least 
partially, within a purely classical framework, provided only that the initial
conditions be chosen as compatible as possible with the laws of
quantum mechanics. This paper follows exactly that line of thought,
being a refinement of that work in that the initial conditions have 
been chosen in a more correct manner, but still remaining within a 
classical description of the system. One may wonder if the price paid
to have this greater accuracy is too high, since we used exact 
wave--functions, which is equivalent to solve the quantum mechanical 
problem. It is not so,  since significant improvements are obtained using no
matter which distribution, even that of eq. (\ref{eq:rhomicrocanonica}) 
which is based upon purely classical considerations. The other 
two distributions have been chosen just to provide the reader with
a comparison.
This is important for all these cases where extracting the wave function 
is too difficult and one is forced to resort to approximations. 
As already pointed out in \cite{hooker97}, it could be worth  exploring in 
some of such situations.

\vskip 0.3cm
\begin{center}
***
\end{center}
\vskip 0.3cm
This work has been supported by a grant of the Italian MURST.
The hospitality offered by the Consorzio RFX is acknowledged.

\newpage

\begin{tabular}{|c|c|c|c|c|c|} \hline
{\em n} & $\alpha_1$ & $\alpha_2 $ & $ \alpha_3$ & $\alpha_{cl}$
        & $ \alpha_{QM}$ \\ \hline
1 & 15/2         & 3.616        & 8/5        & 1        & 9/2 \\
  & (1.67)       & (0.80)       & (0.36)     & (0.22)   &     \\ 
\hline
2 & 210          & 127.6        & 512/5      & 64       & 156 \\
  & (1.35)       & (0.82)       & (0.66)     & (0.41)   &     \\ 
\hline
3 & 1701         & 1350.6       & 5832/5     & 729      & 1377 \\
  & (1.24)       & (0.98)       & (0.85)     & (0.53)   &      \\ 
\hline
4 & 7920         & 7069.7       & 32768/5    & 4096     & 6720 \\
  & (1.18)       & (1.05)       & (0.98)     & (0.61)   &      \\ 
\hline
5 & 53625/2      & 25313        & 25000      & 15625    & 46875/2 \\
  & (1.14)       & (1.08)       & (1.07)     & (0.67)   &         \\ 
\hline
6 & 73710        & 71561        & 373248/5   & 46656    & 65772 \\ 
  & (1.12)       & (1.09)       & (1.13)     & (0.71)   & \\ \hline
10& 1.328 $10^6$ & 1.325 $10^6$ & 1.6 $10^6$ & 1 $10^6$ & 1.2375 
$10^6$ \\ 
  & (1.07)       & (1.07)       & (1.29)     & (0.81)   & \\ \hline
\end{tabular}

\vskip 3cm
 
{\bf Table 1}: polarizability calculated from the different methods
explained in the text. $\alpha_{1, 2, 3}$ from choices 1, 2, 3 
respectively; $\alpha_{cl}$ from eq. (\ref{eq:alfaclassica}); 
$\alpha_{QM}$ quantum  mechanical value from eq. 
(\ref{eq:alfaquantica}). Between parentheses are reported the ratios 
$\alpha / \alpha_{QM} $. $n$ is the principal quantum number.

\end{document}